\documentclass[aip,superscriptaddress,preprint]{revtex4-1}

\usepackage{graphicx}
\usepackage{dcolumn}
\usepackage{bm}

\usepackage[utf8]{inputenc}
\usepackage[T1]{fontenc}
\usepackage{mathptmx}
\usepackage{etoolbox}
\usepackage{amsmath, amsfonts}
\usepackage{tikz}
\usepackage{enumerate}
\usepackage{url}
\usetikzlibrary{shapes.geometric}
\usetikzlibrary{decorations.pathreplacing, angles, quotes}
\usetikzlibrary{arrows, decorations.markings}
\usetikzlibrary{plotmarks}
\usepackage{rotating,array}

\makeatletter
\def\@email#1#2{%
 \endgroup
 \patchcmd{\titleblock@produce}
  {\frontmatter@RRAPformat}
  {\frontmatter@RRAPformat{\produce@RRAP{*#1\href{mailto:#2}{#2}}}\frontmatter@RRAPformat}
  {}{}
}%
\makeatother

\newtheorem{defn}{Definition}

\begin{document}


\title{The Essential Synchronization Backbone Problem}
\author{C. Tyler Diggans}
\email{digganct@clarkson.edu}
\affiliation{Air Force Research Laboratory Information Directorate, Rome , NY 13441, USA}
\affiliation{Department of Physics, Clarkson University, Potsdam, NY 13669, USA}
\affiliation{Clarkson Center for Complex Systems Science (C$^3$S$^2$), Potsdam, NY 13669, USA}

\author{Jeremie Fish}
\author{Abd AlRahman R. AlMomani}
\author{Erik M. Bollt}
\affiliation{Clarkson Center for Complex Systems Science (C$^3$S$^2$), Potsdam, NY 13669, USA}
\affiliation{Department of Electrical and Computer Engineering, Clarkson University, Potsdam, NY 13669, USA}

\date{\today}

\begin{abstract}
Network optimization strategies for the process of synchronization have generally focused on the re-wiring or re-weighting of links in order to: (1) expand the range of coupling strengths that achieve synchronization, (2) expand the basin of attraction for the synchronization manifold, or (3) lower the average time to synchronization. A new optimization goal is proposed in seeking the minimum subset of the edge set of the original network that enables the same essential ability to synchronize in that the synchronization manifolds have conjugate stability. We call this type of minimal spanning subgraph an Essential Synchronization Backbone (ESB) of the original system, and we present two algorithms: one is a strategy for an exhaustive search for a true solution, while the other is a method of approximation for this combinatorial problem. The solution spaces that result from different choices of dynamical systems and coupling schemes vary with the level of hierarchical structure present and also the number of interwoven central cycles. Applications can include the important problem in civil engineering of power grid hardening, where new link creation may be costly, and the defense of certain key links to the functional process may be prioritized.
\end{abstract}

\maketitle

\begin{quotation}
In many applications of dynamical processes, qualitative understanding of long term behavior is often sufficient, while the details leading to that outcome are relatively unimportant. For instance, if a power grid's synchronization is disrupted by the removal of a transmission line, it may be helpful to estimate the time needed to return to synchronization, but more important is the stability of synchronization under the new configuration. We define an optimization problem associated with the core part of the network necessary for synchronization of coupled oscillators, which we call the Essential Synchronization Backbone problem. We seek an edge-minimal spanning subgraph of the original network that preserves the stability of the synchronous states of the original system. The average time to synchronization and the actual basin of attraction may differ wildly, furthermore, adjustment of the global coupling strength is likely required; however, we are satisfied if the new system can achieve the same qualitative type of synchronization. We provide an exhaustive algorithm for obtaining solutions for small systems and a greedy algorithm for approximating solutions to larger systems. Structural trends of hierarchy and loopiness in the solution space depend on the choice of oscillator type and coupling scheme. We consider solutions in the context of linearly diffusive chaotic oscillators because their information theoretic aspects provide insight into the role of graph conductance and hierarchy in the synchronization process.     
\end{quotation}

\section{Introduction}
A common goal in network science is the optimization of network structure for a desired purpose. In the case of synchronization, we do not normally wish to remove any of the dynamical systems (nodes); and so, we are left with the question of how to rearrange links (edges) in the network. Optimal networks for synchronization processes have been widely studied in different contexts, but the strategies employed usually involve re-wiring or re-weighting the edges of an often directed and/or weighted graph~\cite{Restrepo06,Hagberg08,Jalili13,Skardal14,Skardal16,Fazlyab17,Daley20}, and sometimes imposing directionality upon the edges of an initially simple graph~\cite{Zeng09}. Due to potential costs associated with link creation (or imposing directionality for some applications), we instead consider a simple question for synchronization on an undirected and unweighted graph: ``Given a synchronizing network of oscillators, what is the minimal collection of edges that must be retained from the original network for the same type of synchronization to remain possible?" We call any such edge-minimal spanning subgraph of the original network an Essential Synchronization Backbone (ESB) of the system. This concept may prove useful in the assessment of reliability for processes coupled through undirected networks, e.g., the power grid~\cite{Nishikawa15}. 

For the sake of brevity, some general knowledge within the field is assumed, but summaries of the chaotic attractors used and the Master Stability Function formalism are included as appendices for convenience. In general, the ESB of a system is not unique, and furthermore it depends on the dynamical systems at the nodes as well as the chosen coupling scheme, especially for multivariate oscillators. We consider the proposed problem on undirected networks, though it is still well defined in the more general case of directed and/or weighted networks. However, it has been shown that the MSF approach is not always sufficient to characterize the stability of the synchronous state in such cases~\cite{Fish17, Muolo21}, and so any effort to adapt the presented methods must address these issues accordingly. Additionally, for simple and harmonic (i.e. non-chaotic) oscillator systems coupled through undirected and unweighted network architectures, the Essential Synchronization Backbone problem is relatively trivial to solve by the Master Stability Function (MSF) formalism~\cite{Pecora98}; namely, any spanning tree of the network will suffice. This is due to an unbounded interval of values where the MSF is negative, meaning the global coupling strength can always be increased enough to push the normalized spectra into this range. However, this unbounded strength may not be practical in real-life systems. Subsequent iterations of this work may seek to optimize further among these simple solutions, but the present work focuses on chaotic oscillator systems where we find a single bounded interval for which the MSF is negative, indicating linear stability of the synchronization manifold. Under these circumstances, the solutions are not generally trees, but tend to have common network features for a given choice of system, such as levels of hierarchy and the amount of interwoven central loops. This context was chosen due to implications for better understanding the synchronization process as a process of information exchange and transfer~\cite{Stojanovski97,Wang09,Bollt12,Lin14}. 

Since \textit{a picture is worth a thousand words}, we begin with an illustrative example on a small network, referencing the appendices as needed; this is followed by a clear definition of an \textit{Essential Synchronization Backbone}. Then, the chosen context of linear diffusively coupled chaotic oscillators is described in detail, for which solving the ESB problem is approached through two algorithms. The first is an exhaustive search strategy that finds a true ESB of the system; and, the second is a greedy algorithm for approximating an ESB, which enables the study of larger networks. The greedy approach is successful in obtaining a good solution (in terms of a small volume spanning subgraph) with reasonable computational time, but falls short of finding a true ESB in most cases. As such, we will refer to these approximate solutions as Greedy Essential Synchronization Backbones (and denote them as GESB) to clearly indicate the approximation. The details of this discrepancy will be made clear from the initial example.

We also provide preliminary analysis of the solution space as a function of the type of dynamical system and coupling being considered through comparison of two contrasting representative systems. The presence of hierarchical structure and interlocking central cycles lead us to make conjectures about the role of graph conductance (or isoperimetric number) of the network in the process of synchronization through information flow. Chaotic oscillators are particularly interesting in this context since they can be imagined as information generators through symbolic dynamics~\cite{Bollt13}.

\section{The Essential Synchronization Backbone Problem}
Given a network of coupled oscillators that achieve synchronization, what is the edge-minimal spanning subgraph of this network that has similar synchronization behavior? This question is ambiguous in a few ways, and preliminary definitions are required to provide an accurate and succinct problem statement; however, for those familiar with the literature in the field, we first explore a simple example to illustrate the objective clearly and point out certain nuances that are generally encountered.
\subsection{An Introductory Example}
\label{ESBPEx}
Consider a system of $N=12$ identical R\"{o}ssler oscillators (with parameters $a=b=0.2$, $c=9.0$) that exhibit linear diffusive coupling through the $x$-variables over an undirected and unweighted network $G=(V,E)$. Linear diffusive coupling was considered due to the ease with which the Master Stability Function (MSF) formalism can be applied~\cite{Pecora98}; the definitions of some chaotic oscillators and a review of the Master Stability Function are provided in Appendices~\ref{Chaotics} and ~\ref{MSF} respectively for the reader's convenience, and a more detailed definition of a linear diffusively coupled chaotic oscillator system is provided in section~\ref{Background}.

\begin{figure}[ht!]
\centering
\begin{tabular}{cccc}
\includegraphics[width=1.5in]{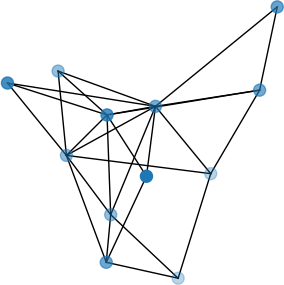}&
\includegraphics[width=1.5in]{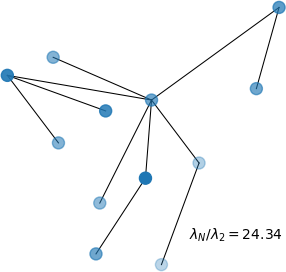}&
\includegraphics[width=1.5in]{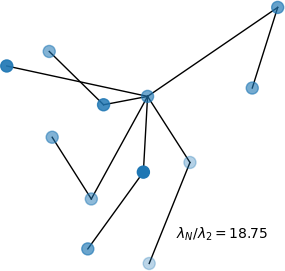}&
\includegraphics[width=1.5in]{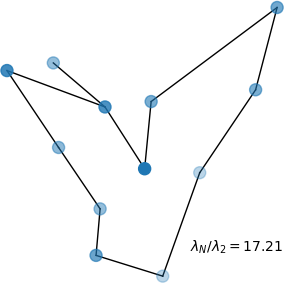}\\
Original & ESB & ESB & GESB\\
(a)&(b)&(c)&(d)\\
\end{tabular}
\caption{(a) The original graph $G$, together with (b) a true Essential Synchronization Backbone (ESB), (c) the minimal eigenratio ESB, and (d) a greedy approximation of an ESB (GESB) for the system of identical R\"{o}ssler attractors using $x-x$ coupling on the network $G$.\label{Graphs}}
\end{figure}

For our immediate purposes, it was shown that the MSF for the above system is negative on a single interval of coupling-normalized eigenvalues~\cite{Huang09} ($K\in[K_\alpha, K_\beta]$, where $K_\alpha\approx 0.186$ and $K_\beta\approx 4.614$). For example, we take $G$ to be the graph whose embedding is shown in Fig.~\ref{Graphs} (a) and we denote the ordered eigenvalues of the combinatorial graph Laplacian associated with $G$ as $0=\lambda_1<\lambda_2\leq\cdots\leq\lambda_N$. In this case, the synchronization manifold of the system is exponentially asymptotically stable for an appropriately chosen global coupling strength due to the eigenratio of $R=\lambda_N/\lambda_2 \approx 9.15$ for this particular choice of $G$ being less than the synchronizability ratio of the system ($R_{MSF}=K_\beta/K_\alpha\approx 24.8$).

Our goal centers on the question of whether we can remove an edge from the network $G$ to obtain a spanning subgraph $G'=(V,E')$ (where $E'\subseteq E$) such that $R'=\lambda'_N/\lambda'_2 < R_{MSF}$ as well, where the prime indicates association with the subgraph $G'$. We are not interested in how the global coupling strength may need adjustment nor the average time to synchronization, just whether synchronization can occur. If such a spanning subgraph exists, then the invariant manifolds of the two systems are considered conjugate, and this motivates our use of the adjective \textit{essential}, meaning ``in essence".
  
If one edge can be removed in this way, can another? This begs the most obvious next question: What is the maximum number of edges that can be removed from the network that results in a system with similar synchronization behavior? While the incremental approach outlined is not successful for achieving our aim in general, it is the basis for our approximation method. Deciding which edge(s) to remove is nontrivial, especially for large networks where one may be able to remove almost any edge initially. One could exhaustively check the impact on the eigenratio $R'$ for all possible edge removals at each step in the process, removing the edge that results in the lowest $R'$ value for that step (we refer to this strategy as the hybrid approach since it is an exhaustive search for making greedy decisions), but it is rather computationally expensive.

A useful observation was made in \textit{Hagberg et. al. (2008)}~\cite{Hagberg08} about the impact of edge removals on the spectrum of the graph Laplacian: considering the eigenvector corresponding to $\lambda_N$, the edge with the largest difference in valuation on its boundary is a good candidate for removal for the goal of minimizing the eigenratio~\cite{Hagberg08}. This heuristic has been used for the goal of reducing redundant links, while maintaining most of the original synchronization process's properties~\cite{Zhang14}. For our simpler goal of conjugate stability, the identified link is not always optimal due to the added normalization requirement, but making the optimal short term choice may already prohibit finding the global solution, so we let this guide a greedy algorithmic approach that is very computationally efficient and still obtains good approximations for large networks. In this way, we can find spanning subgraphs of very large graphs that also have exponentially asymptotically stable synchronization manifolds, even if they are not truly the minimal-edge case. 

To illustrate, Fig.~\ref{Graphs} (b) shows a true Essential Synchronization Backbone (ESB) for the system described above coupled through the network shown in Fig.~\ref{Graphs}(a). The ESB shown in Fig.~\ref{Graphs} (c) is also a true ESB of the system, but it is what might be referred to as a \textit{minimal eigenratio ESB} for the system. However, for our purposes, these ESB are just two of $1899$ equivalent solutions for this networked oscillator system. It is also informative to point out that there are $553,521$ other spanning subgraphs of $G$ with the same number of edges (spanning trees in this case) that are not ESB of this system, having eigenratios larger than the system's $R_{MSF}$. As may be apparent by the size of the search space for such a small network, the exhaustive search of the power set of the edge set is not feasible for larger networks, and so alternative algorithms for approximating the ESB are of interest.

Figure~\ref{Graphs} (d) shows the result of our greedy approximation algorithm (which is available in \textit{python} at \url{https://github.com/tylerdiggans/ESBP}) for this system. This approximate solution has an additional edge, meaning it is not actually an ESB of the system. We call this type of local minima, for which no further single edge can be removed a Greedy-ESB or GESB to differentiate it from a true solution. However, it is worth pointing out that this approximation has a smaller synchronizability ratio than the minimal eigenratio ESB, meaning its structure is more easily synchronized in general. This points to a limitation of our algorithmic approach being focused on the minimization of the eigenratio at each step, since this property is non-monotonic with edge removals~\cite{Ravoori11} and is not representative of our actual goal. In summary, this example aims to convey: 
\begin{itemize}
\item The ESB is not unique, e.g., Fig.~\ref{Graphs} (b) and (c) are both ESB of the system on $G$ (Fig.~\ref{Graphs} (a))
\item Optimization with respect to other objectives is possible within the (nontrivial) set of ESB
\item Our goal is minimization of $|E'|$ such that $R'<R_{MSF}$, but $R'$ itself need not be minimal
\item The Greedy approach often results in local minima due to over-optimizing on $R'$
\end{itemize}

While the above example seems to indicate that the GESB are not representative of the general structure of the ESB, this is an artifact of our intentional choice to illustrate where this approximation breaks down. While all ESB for this small network are spanning trees, this is due to the small number of nodes. The general trends in structure of ESB vary greatly depending on the type of dynamical systems being synchronized and the choice of coupling (e.g., see Fig.~\ref{DR12_4} for an example). The ESB for systems where the MSF is negative on a half-open interval $[K,\infty)$ are always spanning trees due to having no restriction on the coupling strength. This is the case for simple and harmonic oscillators, as well as some choices of coupling for chaotic oscillators. But, for instance, when the MSF is only negative on a closed interval, or if a restriction on the global coupling strength prevents the utilization of the unbounded interval, we find more interesting structure that seems related to the required information flow to maintain synchronization for such oscillators with positive topological entropy~\cite{Bollt13}, where predictable oscillations are not present. We will explore these details toward the end of this paper and in future work. For now, we take this space to carefully define the Essential Synchronization Backbone.

\subsection{An Optimization Problem}
\label{ESBPDefn}
We begin with a constructive definition that establishes the meaning of the term \textit{essential} within this context.
\begin{defn}
\label{same}
Let $A$ and $B$ be two networked dynamical systems sharing a domain $\Omega\subseteq\mathbb{R}^{M}$. 
\begin{enumerate}[a)]
\item The invariant manifolds of systems $A$ and $B$ are \textbf{essentially equivalent} if there exists two sets $U, V\subseteq \Omega$ such that $A|_U$ and $B|_V$ are conjugate. 

\item Further, if the invariant manifolds of systems $A$ and $B$ are stable, we say they have \textbf{essentially equivalent stability}. 
\end{enumerate}
\end{defn}

Recall the definition of a spanning subgraph of a connected simple graph~\cite{Nica18}:
\begin{defn}
The network $H=(V,F)$ with vertex set $V$ and edge set $F$ is called a \textbf{spanning subgraph} of the simple network $G=(V,E)$ if: (1) $H$ is connected and (2) $F\subseteq E$ , i.e. $H$ is obtained from $G$ by edge removals. 
\end{defn}

Given the above definitions, we now define an Essential Synchronization Backbone (ESB) of a networked oscillator system:
\begin{defn} 
Let $A$ and $B$ be two systems, each consisting of $N$ networked oscillators that share a particular choice of uncoupled dynamics $f$ and coupling scheme, but interact through the networks $G$ and $G'$ respectively. Given system $A$ has a stable synchronization manifold, system $B$ is called an \textbf{Essential Synchronization Backbone} (ESB) of system $A$, if $G'$ is a minimal-edge spanning subgraph of $G$ such that systems $A$ and $B$ have \textrm{essentially equivalent stability}. 
\end{defn}

\section{Background}
The study of synchronizing oscillators often focuses on simple or harmonic oscillators with complexity being introduced through nonlinear coupling, e.g. the Kuramoto model. But, the study of chaotic oscillators adds an alternate type of complexity that is of particular interest to the study of information flow. As such, we have explored the ESB problem within the context of linearly diffusive coupling of identical chaotic oscillators over unweighted and undirected networks, for which exponential asymptotic stability of the synchronization manifold is fully characterized by the MSF~\cite{Pecora98}. In the more general context of directed networks, recent work has suggested that the MSF may not fully characterize the stability of the synchronous states~\cite{Fish17, Asslani18, Muolo21}, but this point is contentious\cite{nishikawa2021comment, muolo2021reply, sorrentino2022comment, fish2022nonnormality}; regardless, we restrict our discussion to the undirected case. The present choice of dynamics was made in order to focus attention on the interesting results that arise in this context, but the general definitions above apply equally to synchronization processes with simple, harmonic, or nonidentical oscillators (where Lyapunov stability would be the relevant type~\cite{Sun09}). 

\subsection{Networked Chaotic Oscillators}
\label{Background}
We consider a collection of $N$ identical oscillators, which in isolation would follow the differential equation, $\dot{\textbf{x}} = f(\textbf{x})$, referred to as the uncoupled dynamics. Linear coupling is a good local approximation to nonlinear schemes through Taylor's approximation, e.g., $x\approx \sin(x)$; so, we choose to study linear diffusive coupling.

We let $\vec{X}=[\textbf{x}_1,\textbf{x}_2,...,\textbf{x}_N]\in\mathbb{R}^{M}$ be the set of dynamical variables of the system consisting of $N$ identical dynamical oscillators, each of dimension $d$, i.e. $M=Nd$. These dynamical systems are connected by an unweighted and undirected network $G=(V,E)$. The component variables are governed by the uncoupled dynamics $\dot{\textbf{x}}=f(\textbf{x})$, together with linear diffusive coupling through both a linear coupling function, $H(\vec{X})= [\hat{H}\cdot \textbf{x}_1, \hat{H}\cdot \textbf{x}_2, \dots,\hat{H}\cdot \textbf{x}_N]$ with $\hat{H}\in\mathbb{R}^{d\times d}$, and the network topology, represented by the combinatorial graph Laplacian associated with $G$, which we denote by $L$. Then, for a specified global coupling strength $\sigma$, the composite variable $\vec{X}$ follows the dynamics
\begin{equation}
\dot{\vec{X}}=F(\vec{X})- \sigma L\otimes H(\vec{X}),
\label{x}
\end{equation}
where $F(\vec{X})=[f(\textbf{x}_1), f(\textbf{x}_2),... ,f(\textbf{x}_N)]$ represents the uncoupled dynamics.  

To contrast the results given for the R\"{o}ssler oscillators described above, which we will refer to going forward as a R\"{o}ssler-$x$ system, we also consider Lorenz oscillators coupled through their $z$ components (Lorenz-$z$). Both oscillators are defined in Appendix~\ref{Chaotics}, and the MSF for both system choices are included in Appendix~\ref{MSF}. The Lorenz-$z$ system technically results in an MSF of the third type, meaning it has both a finite and an infinite interval of coupling-normalized eigenvalues where the MSF is negative~\cite{Huang09}. Since this would indicate that all ESB would be spanning trees, we have decided to impose a large but finite restriction on the global coupling strength. The infinite interval is sufficiently far from the finite one that this effectively restricts the MSF to being negative only on the finite interval $[1.422,6.035]$, implying $R_{MSF}\approx 4.24$ for this modified Lorenz-$z$ system. This type of restriction is reasonable in physical systems, so not altogether artificial, especially when interested in avoiding the scenario referred to as \textit{dominance of neighbors}~\cite{Sun2014}, where the coupling influence dominates the internal dynamics. This modification enables the consideration of familiar oscillators with simple coupling; additionally, the particular solution space to the ESBP for this smaller $R_{MSF}$ value is instructive for comparison. Limiting our discussion to these two three-dimensional chaotic oscillators, we only require two choices of the matrix $\hat{H}$, and so we define them explicitly as:
\begin{equation}
\hat{H}_x = \left[\begin{array}{ccc}
1 & 0 & 0\\
0 & 0 & 0\\
0 & 0 & 0
\end{array}\right] \hspace{25pt}  \textrm{and} \hspace{25pt}  \hat{H}_z = \left[\begin{array}{ccc}
0 & 0 & 0\\
0 & 0 & 0\\
0 & 0 & 1
\end{array}\right],
\end{equation}
meaning, R\"{o}ssler-$x$ systems use $\hat{H}_x$, while Lorenz-$z$ systems use $\hat{H}_z$. While we focus on these two particular choices of dynamical system and coupling, they are representative of any coupled system choices with comparable MSF structures. Since, in these cases, the $R_{MSF}$ value completely determines the spectral properties of all networks that will synchronize, any two systems with sufficiently different $R_{MSF}$ values will lead to such contrasting results for the ESB problem.

Again, the general concept of an ESB may be interesting in other contexts, e.g. heterogeneous dynamical systems, where the synchronization manifold may only be Lyapunov stable, but for the case of identical oscillators, due to the form of $L$, the \textit{synchronous states} take the form $\vec{X} = [\textbf{x}^*,..., \textbf{x}^*]$ where $\textbf{x}^*$ is a solution to the uncoupled dynamics $\dot{\textbf{x}}=f(\textbf{x})$. The set of all such states make up the synchronization manifold, and the type of associated stability in this case is exponentially asymptotic stability, which can be determined in the case of undirected networks using the MSF (see Appendix~\ref{MSF} for details).

\subsection{Optimal Synchronizability}
Over the years, some general trends and heuristics have been identified for enhancing synchronizability over undirected networks, but these observations have usually led to common misconceptions such as:
\begin{itemize}
\item Minimizing average path lengths is important to synchronization. This is true to some extent, and is assumed to be related to the efficiency of information flow; but this trend is not universal as was shown for scale-free networks~\cite{Nishikawa03} where heterogeneity in the connectivity led to the destruction of synchronization while also leading to shorter path lengths. 
\item The bottleneck to synchronization may be the hub nodes and/or links with the largest load~\cite{Belykh05}, but this is highly dependent on the type of dynamics and behavior of synchronization for star graphs.
\item Also, synchronization can be destroyed by a single well-placed link~\cite{Belykh05}, meaning it may really be the weakest link that is the major concern (we recall Braess' Paradox~\cite{Braess05,Witthaut12})
\end{itemize}

These properties accurately describe many features of small world networks, and this class has been explicitly considered in the literature~\cite{Barahona02,Pecora08}. While these may not be true in every case, they still provide valuable insight into common properties of good networks for synchronization. Overall, what is generally sought is a network with relatively small average path length, hubs that are not too overloaded, and avoidance of any catastrophic edges. Quantifying all of these properties simultaneously is challenging and we see the concept of synchronization as information exchange and transfer~\cite{Bollt12} as the proper framework to search for answers.

Given the system~(\ref{x}) and the MSF formalism for determining stability of the synchronization manifold, the problem of finding an Essential Synchronization Backbone of a network $G=(V,E)$ for which $R\leq R_{MSF}$ (given a choice of dynamics and coupling), becomes the question of finding a minimal-edge spanning subgraph $G'=(V,E')$ such that $R'=\lambda'_N/\lambda'_2 < R_{MSF}$ as well, where again, we have used the prime notation to indicate variables and spectra associated with the subgraphs. Thus, we are most interested in how edge removals will affect the two particular eigenvalues of $\lambda_2$ and $\lambda_N$ for the Laplacian of the original network due to their role in the MSF analysis. 

It is known that edge removals will not increase $\lambda_2$ due to interlacing eigenvalues; and in fact, by Weyl's inequality for the spanning subgraph, we know that $\lambda_2-2\leq\lambda'_2\leq\lambda_2$ when one edge is removed from $E$. More generally at any stage in the sparsification, we know that $\lambda'_2\geq\frac{4}{N D}$, where $D$ represents the diameter of the spanning subgraph~\cite{Mohar91}. So, while the more relevant eigenvalue in relation to the effects of edge removal is $\lambda_N$, these lower bounds on $\lambda_2$ are relevant to understanding the patterns of cycles present in ESB. 

The main results on the impact of edge removals from \textit{Restrepo et. al., (2006)}~\cite{Restrepo06} (formulated for directed networks) also apply to undirected graphs; however, that treatment relies on ignoring second order perturbations, which may not remain relevant as the perturbations become large with respect to the size of the network. 

Instead, we look to \textit{Hagberg et. al., (2008)}~\cite{Hagberg08} for guidance on algorithm development based on incremental edge removal, where it was noted that the largest eigenvalue is characterized by
\begin{equation}
\begin{aligned}
\lambda_N &= \sum_{i\sim j}{\left(v_{N}(i)-v_N(j)\right)^2}\\
& \textrm{where} \sum_i v_N(i)^2=1,
\end{aligned}
\label{eigensum}
\end{equation}
where $v_N$ is the eigenvector associated with $\lambda_N$. Thus, in order to reduce $\lambda_N$, one seeks to remove the edge that contributes the largest amount to the constrained sum~(\ref{eigensum}). While the normalization requirement makes this term nontrivial to identify, generally, removing the edge with the largest difference in $v_N$ reduces $\lambda_N$~\cite{Hagberg08}. We now present two methods for solving the ESBP, the second of which relies heavily upon this heuristic.

\section{Solving the ESBP}
Solving the ESBP is an NP-hard problem due to the combinatorial search over the edge set of the original network. Furthermore, for each choice of edge removal, an $O(N^3)$ algorithm is required to test the MSF eigenratio of each solution, meaning brute force approaches to solving this problem are only reasonable for very small and/or sparse networks ($N\sim 10$). Regardless, a parallelized algorithm can explore all potential ESB graphs in an organized manner for small graphs, though considerable resources are required for even sizes of $N\sim 10$. We present this brute force approach, followed by an efficient algorithm for obtaining greedy approximations of solutions (GESB) for larger complex networks.

\subsection{The Exhaustive ESB Algorithm}
Since the spanning subgraph must be connected in order to synchronize, we begin an exhaustive search by considering all sets $E'\subseteq E$, such that $|E'|=N-1$, selected from the edge set of the original graph $G=(V,E)$. Many of these subsets will not result in a connected graph and are discarded; those that result in a connected subgraph are by definition spanning trees. Comparing the eigenratio $R'=\lambda'_N/\lambda'_2$ for the Laplacian of these spanning subgraphs $G'=(V,E')$ with $R_{MSF}=K_\beta/K_\alpha$ from the MSF, it is quickly determined whether the synchronization manifolds of any of these spanning subgraphs and the original network have \textit{essentially equivalent stability}, i.e. both $R$ and $R'$ are less than $R_{MSF}$. If all such subsets of size $N-1$ edges have been considered without finding any that pass this MSF comparison, we increment the number of selected edges and repeat the process until such a graph is found with $R'\leq R_{MSF}$. 

Due to the order in which the spanning subgraphs are considered, as soon as any $G'$ satisfies the condition, that graph is guaranteed to be an ESB of the network, though there is no expectation that the ESB should be unique. In order for all ESB of the system to be found, one must complete comparisons for all subgraphs with the same number of edges as $E'$, and if interested in further optimization, one might compare the eigenratios of these equivalent volume solutions to obtain a \textit{minimal eigenratio ESB}. In the worst case, we find that $E'=E$ is the first subset for which $R'<R_{MSF}$, and our network $G$ is its own backbone for the system in question.

This basic algorithm (available at \url{https://github.com/tylerdiggans/ESBP}) is not practical for most networks, and its performance varies greatly with the system considered due to the higher expected number of edges in an ESB for systems with smaller $R_{MSF}$ values. However, it provides a reasonable method of obtaining the true ESB in order to validate less expensive approximation algorithms. Utilizing parallel computation and efficient approximations of $\lambda'_2$ and $\lambda'_N$ by power methods can help speed up the process, but limitations are encountered quickly for graphs with $N>10$ nodes.

\subsection{Greedy-ESB Algorithm}
\label{GESB}
Since most applications of interest would have hundreds of nodes, if not more, the exhaustive search is infeasible for systems on such large complex networks, meaning we are interested in approximating solutions with what we call Greedy-ESB, or GESB. For this purpose, we utilize the results from \textit{Hagberg et. al., (2008)} to guide an iterative greedy choice of edge removals (code for implementation in python is available at \url{https://github.com/tylerdiggans/ESBP}). The process is as follows:

A network is initiated as a copy of the original graph, namely $G'=(V,E')$, with $E'=E$. Then, at each step in the sparsification process, the eigenvector associated with the largest eigenvalue of the current $G'$ is computed and the remaining edges in $E'$ are ordered with respect to their contributions to the sum that defines $\lambda_N$, Eq.~(\ref{eigensum}). Beginning with the edge having the largest contribution and moving down the list one at a time, the eigenratio $R'$ is computed for the network that would result from the removal of that edge until a spanning subgraph is found such that $R'\leq R_{MSF}$. The first edge $e$ to pass this condition is removed from the edge set permanently, i.e. $E'=E'-{e}$, and the process is repeated until no such edge can be found. This new eigenratio need \textit{not} satisfy $R'<R$, as we are only interested in essentially equivalent stability.

Unfortunately, the incremental edge removal process is known to be non-monotonic in terms of the spectrum spread~\cite{Ravoori11}, thus it is likely that this process will reach a local minimum. Additionally, using the eigenvector to guide our choice is not always optimal as the normalization process must be considered to better order the edges for consideration. Although there are lower bounds for $\lambda_2$, it is also possible that edge removals may lead to $\lambda'_2<\lambda_2$, which may increase $R'$. As stated previously, it is possible to check the eigenratio for each edge in each step, choosing the locally optimal edge to remove, but due to the non-monotonicity, such a hybrid approach only performs marginally better than our greedy approach and at a steep cost in computational time (e.g., approximately 50$\times$ for $N=50$). It has also been suggested to use simulated annealing with such approaches to avoid local minima~\cite{Jalili09}, but it has been found that this only marginally improves results, again, at a computational cost. Since we are not actually finding the true ESB in any of these cases, we chose to present the most basic, yet most efficient, algorithm here for brevity, and show that these GESB are still informative of general structure. 

\subsection{Verification of Greedy Algorithm}
Having presented an efficient method of approximation, we wish to verify that the GESB obtained are in fact ``good" approximations. Due to the high computational complexity of the exhaustive approach, this comparison can only be carried out for relatively small order networks, meaning any verification provided in this domain may not generalize well to more complex networks. Regardless, from this initial comparison, it should be assumed that all GESB are not true ESB, but that they are still reflective of general trends in hierarchical structure and loopiness.

\begin{figure}[ht!]
\centering
\includegraphics[width=0.9\textwidth]{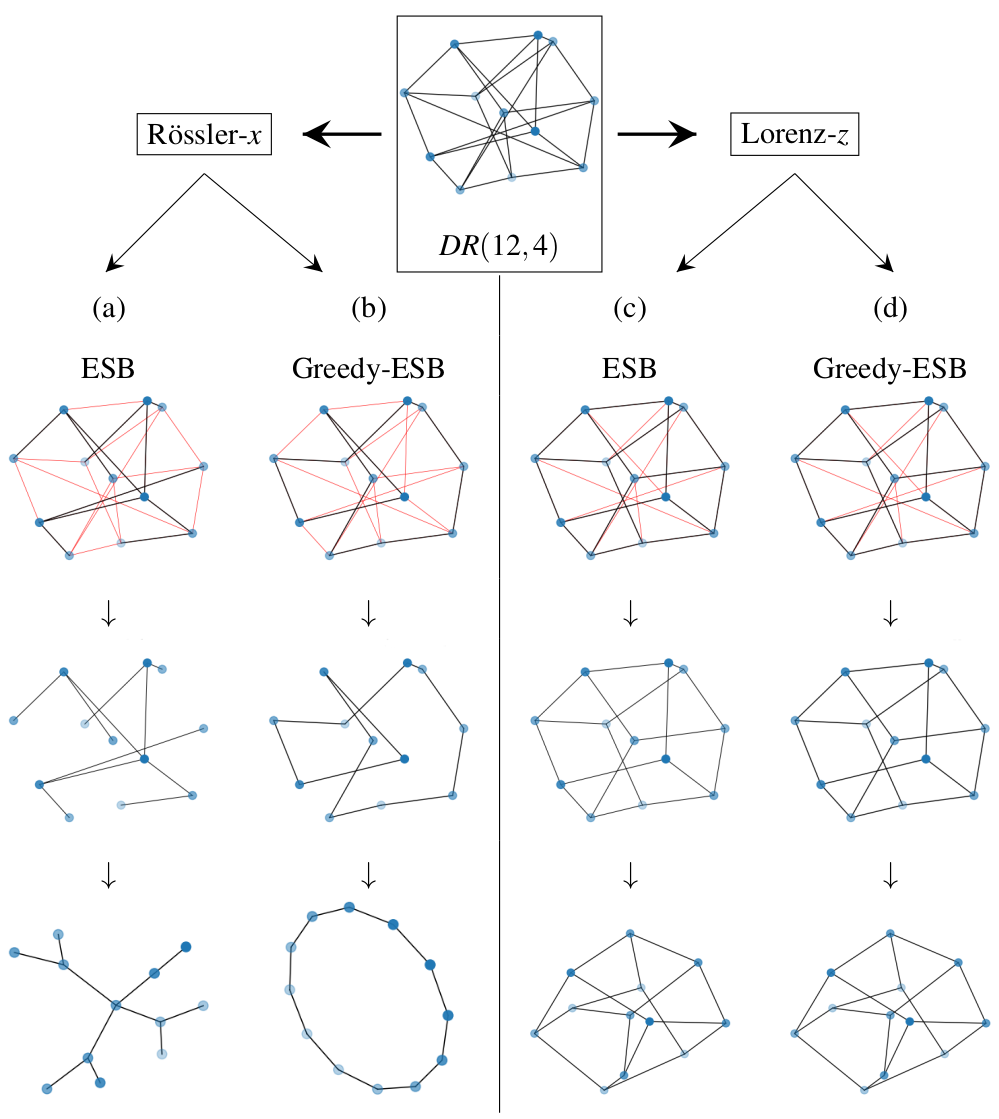}\\
\caption{An example showing the variation in ESB and GESB for different choices of dynamics and coupling on the graph $DR(12,4)$, which is a $d$-regular graph with $N=12$ vertices and $d=4$. Synchronizable for both R\"{o}ssler-$x$ and modified Lorenz-$z$ systems, each column shows the backbone in black on the original graph with the removed edges as thin red lines, followed by the backbone itself, and finally an alternate embedding that showcases the backbones structure; (a) and (b) show the ESB and GESB resp. for the R\"{o}ssler-$x$ system, (c) and (d) show the ESB and GESB resp. for the Lorenz-$z$ system. Note that both algorithms find the same ESB in the case of the Lorenz-$z$ system, although this is not true in general. \label{DR12_4}}
\end{figure}

We begin by considering a $d$-regular graph of $N=12$ vertices with $d=4$, which we denote by $DR(12,4)$. This network provides a stable synchronization manifold for both R\"{o}ssler-$x$ and modified Lorenz-$z$ type systems, and so it illustrates the impact of the chosen dynamics and coupling on the overall results. For each choice of dynamics and algorithmic solution, a series of plots is provided in Fig.~\ref{DR12_4}, which in descending order are as follows: the original graph with the removed edges highlighted in red, the backbone in the original graph embedding, and the backbone shown in a new embedding to showcase the structural trends for each solution. Figure~\ref{DR12_4} (a) and (b) show the sequence of plots for a true solution for the R\"{o}ssler-$x$ system and our Greedy search respectively. Figure~\ref{DR12_4} (c) and (d) show the sequence of plots for a true solution to the Lorenz-$z$ system and our Greedy search respectively. 

\begin{figure}[ht!]
\centering
\includegraphics[width=0.925\textwidth]{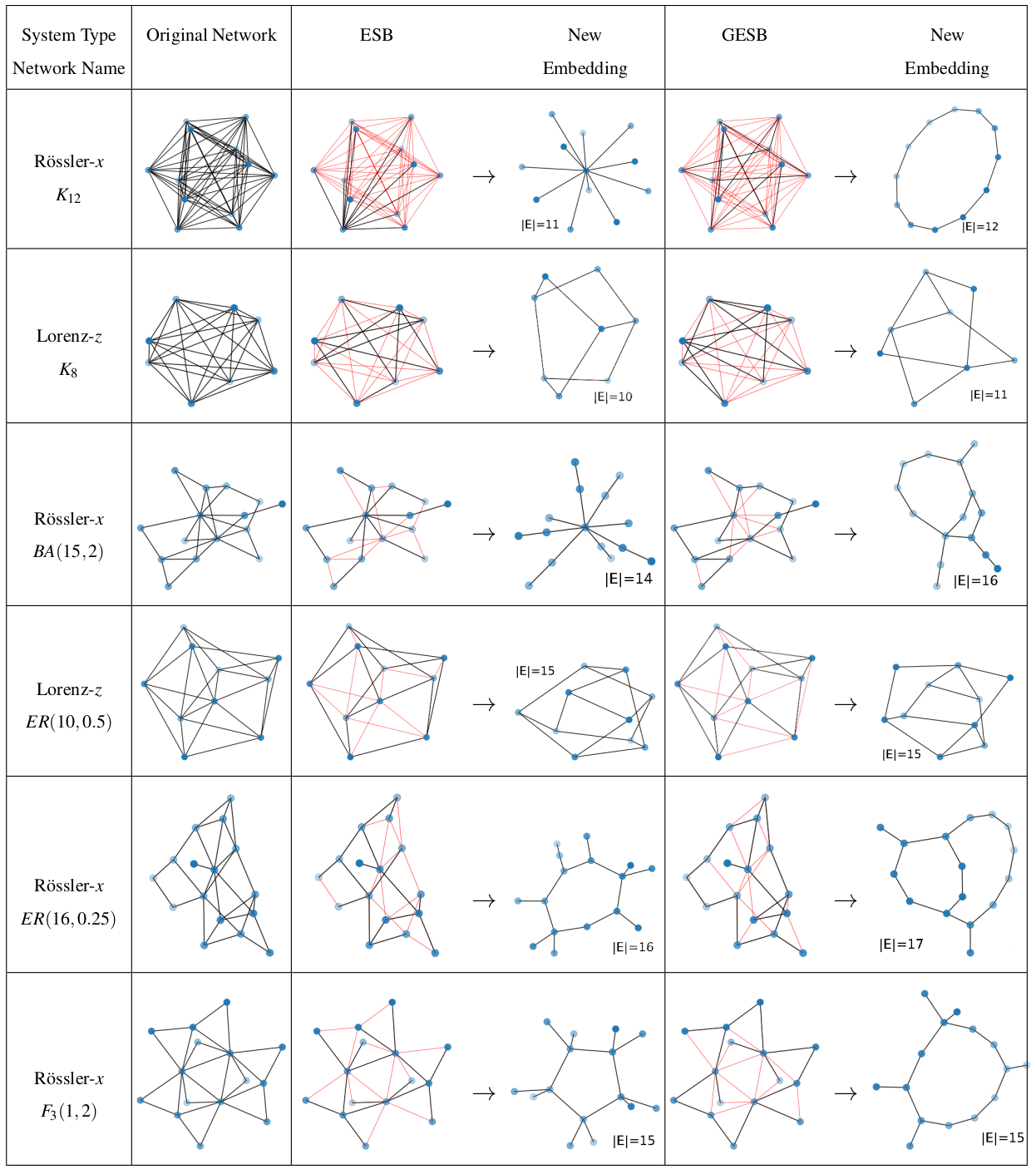}
\caption{Representative networks considered for verification of the greedy algorithms performance, together with their true ESB and Greedy-ESB obtained from the brute force and greedy algorithms resp. Note that the Greedy algorithm often results in local minima having at least one extra edge, though not always. $K_n$ represents the complete graph on $n$ vertices; $BA(N,m)$ represents a Barab\'{a}si-Albert graph~\cite{Barabasi99} with $N$ vertices and growth parameter $m$; $ER(N,p)$ represents an Erd\"{o}s-Reny\'{i} graph on $N$ vertices with parameter $p$; and $F_n(u,v)$ represents the $n^{\textrm{th}}$ generation of a $(u,v)$-flower graph~\cite{Rozenfeld06}.\label{Gallery}}
\end{figure}

To show general trends more clearly, a gallery of plots for our two choices of system interacting on various networks is provided as Fig.~\ref{Gallery}. Here, in each case, we provide the original network, and both a true ESB of the system and the associated GESB obtained by our greedy algorithm, where we include the solutions in the original embedding with removed edges highlighted in red, followed by an updated embedding that highlights the structure. 

\begin{figure}[ht!]
\centering
\includegraphics[width=0.9\textwidth]{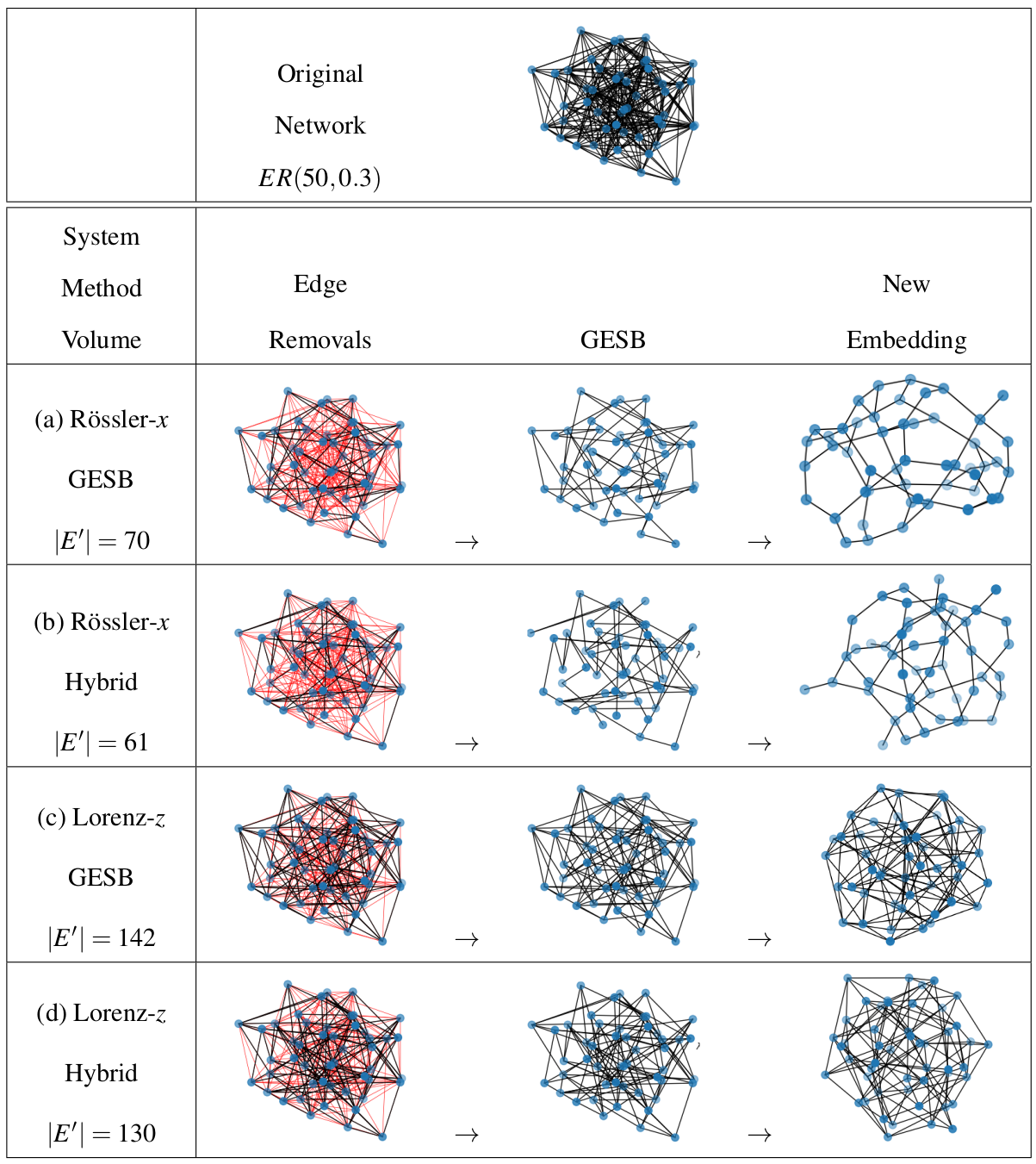}
\caption{An Erd\"{o}s-Reny\'{i} graph ($ER(50,0.3)$) having $|E|=345$ that is synchronizable for both choices of dynamics; (a)-(d) provide a collection of Greedy-ESB (GESB), obtained by two different algorithmic approaches for the two chosen systems: (a) and (b) are GESB for the R\"{o}ssler-$x$ system obtained using the basic greedy algorithm and the hybrid approach (described in section~\ref{GESB}) respectively, while (c) and (d) are for the Lorenz-$z$ system. Although the hybrid approach usually finds smaller volume spanning subgraphs (better approximations), the basic greedy approach is more computationally efficient and both procedures result in GESB having similar structure for a given type of oscillator system. \label{LargerOnes}}
\end{figure}

It is also instructive to consider the GESB obtained for larger graphs, although the exhaustive search is infeasible; for comparison, we provide the resulting GESB from the greedy algorithm with the GESB obtained by the hybrid approach described in section~\ref{GESB}. While it is difficult to find large networks for which the modified Lorenz-$z$ system synchronizes at all, an $ER(50,0.3)$ graph with $345$ edges was found that synchronizes for both the R\"{o}ssler-$x$ and modified Lorenz-$z$ systems. Greedy-ESB were approximated for the two systems and are shown in Fig.~\ref{LargerOnes} (a) and (c), along with the volume of their edgesets, and for comparison, the hybrid approach was used to obtain a more optimal GESB, which are shown in Fig.~\ref{LargerOnes} (b) and (d). 

From Fig.~\ref{LargerOnes} (a), the R\"{o}ssler-$x$ GESB had $|E'|=70$ edges, while the hybrid approximation (b) was able to obtain a GESB with $|E'|=61$ edges; (c) the basic GESB for the Lorenz-$z$ system was found to have $|E'|=142$ edges, yet the hybrid approach (d) found a GESB with $|E'|=130$ edges. This indicates that some of our heuristic edge choices were not optimal. We do not expect these hybrid GESB to be true ESB in either case, and instead point out the similarity in large scale features (e.g., the presence of pendant vertices for R\"{o}ssler-$x$ systems) as indicative of success. 

\section{Analysis}
The general result is that while the greedy algorithm finds substantially low volume spanning subgraphs, it is common for the GESB to not be a true ESB of the network. However, as can be seen in the general trends for different choices of dynamical systems, the apparent concern from our original example should be allayed. We find that true ESB for networks that are too large to allow spanning trees result in similar structures to the GESB of our original example. Small spanning trees may be permitted in certain cases such as R\"{o}ssler-$x$ systems, but star networks, which have the optimal eigenratio for any tree, were shown to have size limits on synchronization for such systems~\cite{Pecora98}. Since $\lambda_2\geq \frac{4}{ND}$, as the number of nodes increases, the growing diameter of spanning trees eventually places pressure on the formation of cycles to alleviate the rising eigenratio. First, loops appear in the ESB, which may enable feedback to bolster synchronization (also reducing the diameter, which places a tighter lower bound on $\lambda_2$, thereby preventing the eigenratio from growing too large). 

This trend is followed by the formation of larger small world networks consisting of multiple loops that intersect. We better quantify this trend through analysis of common graph structures such as star and cycle graphs, for which general trends can be immediately identified based on the type of oscillators used. This is further evidence that the amount of information required to synchronize different systems may play an important role in determining a required isoperimetric number for synchronization associated with the number (and interaction) of loops, in the ESB.

Although identifying general trends provides some intuition and can guide experiments, we also need to identify clear limitations of the role that various common network structures and motifs may play in forming a backbone of a more complex network. From the beginning of this work, we have known that there will be limits to spanning trees being ESB of systems of chaotic attractors. In particular, it has been shown that for R\"{o}ssler attractors coupled through $\hat{H}_x$, the star topology will not synchronize for more than $N=45$ nodes~\cite{Pecora98}; related to this is the upper bound of $\lambda_N\leq 2\Delta$, where $\Delta$ is the maximum degree of the network. Similar arguments can be made related to the limitations of spanning trees in general and the bounds for $\lambda_2$. Such trends indicate that for any choice of minimal topology, there will be a limit to synchronization. We now consider a few relevant basic graph topologies to explore these trends further.

When considering synchronization of R\"{o}ssler-$x$ systems, we find that on the cycle graph on $N$ vertices, $C_N$, the system's ESB is a path graph, $P_{N-1}$, for $N<9$, but for $9\leq N\leq 17$, $C_N$ is its own backbone, meaning it is already the minimal-edge subgraph. For $N>17$ the original graph no longer synchronizes at all for this system, which is indicative that we will likely not find such large cycles forming the basis of ESB in larger graphs for this choice of dynamical system and coupling. As can be seen in the GESB in Fig.~\ref{LargerOnes}(b), while there are many interconnected cycles, no element of a cycle basis has more than $17$ nodes included.

Furthermore, the synchronization of Lorenz-$z$ attractors in cycles is more restricted as well. We find that only $C_3$ has an ESB of $P_2$, which may explain why we do not find pendant vertices in the ESB of larger networks for these systems. Whereas $C_N$ is its own backbone for $4\leq N\leq 6$, and then $C_N$ no longer synchronizes for $N>6$. This is further evidence that different dynamical systems require different amounts of information flow to support synchronization, which will be explored in depth in future work.

\begin{figure}[ht!]
\centering
\includegraphics[width=5.5in]{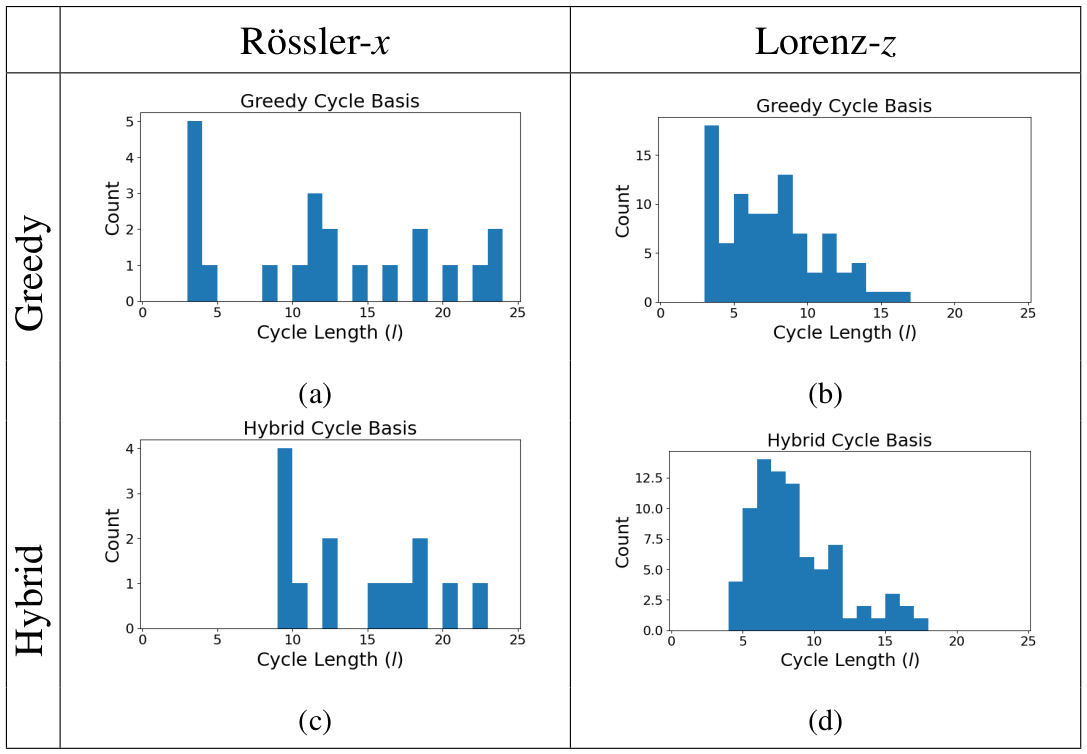}
\caption{Cycle length histograms for the GESB obtained using the Greedy (a) and (b) and Hybrid (c) and (d) approaches on the $ER(50,0.3)$ graph (see Fig.\ref{LargerOnes}), indicating that the limitations of synchronizing $C_n$ are relevant to larger network (the most abundant cycle lengths being $l=9$ and $l=6$ for the R\"{o}ssler-$x$ and Lorenz-$z$ systems resp.). Both algorithms produce similar statistics based on the choice of dynamics, except for the presence of clustering in those found using our greedy algorithm; this indicates the potential for algorithm improvement. \label{Cycles}}
\end{figure}

These limits in cycles are represented in our results, e.g., for networks of Lorenz attractors we find backbones consisting of many intersecting small cycles, but none of the basic cyclic paths is much larger than six edges. A flaw in our greedy approach is illustrated by computing a cycle basis for each GESB in Fig.~\ref{LargerOnes}, for which histograms of the cycle lengths are provided in Fig.~\ref{Cycles}. As expected, the Lorenz-$z$ GESB has many more edges than the R\"{o}ssler-$x$ system, and thus had many more cycles, though they tended toward smaller cycle lengths ($l$) with the most common cycle length of $l=6$ for the more optimal GESB (obtained by the Hybrid approach). This is in contrast to the most common cycle length of $l=9$ for the R\"{o}ssler system. These lengths are consistent with the analysis of $C_N$ above. While additional cycles are found with longer lengths than the restrictions found in $C_N$, it is reasonable to expect that these smaller cycles are enabling the longer ones to synchronize through feedback. Considering the cycle histograms for the GESB obtained by our greedy algorithm indicates a problem: the prevalence of clustering. By comparing the cycle lengths, we can see that the more optimal GESB have zero clustering in both cases, and it makes sense that if cycles are present, in order to reap the rewards of the additional edge, one would want to maximize the length of these cycles within the bounds of synchronization. Future work may seek to remedy this by penalizing clustering in some way.

\begin{figure}[ht!]
\centering
\begin{tabular}{>{\centering\arraybackslash}m{110pt}>{\centering\arraybackslash}m{110pt}>{\centering\arraybackslash}m{110pt}>{\centering\arraybackslash}m{110pt}}
\includegraphics[width=1.25in]{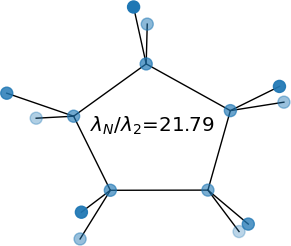}&
\includegraphics[width=1.25in]{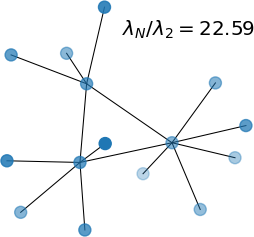}&
\includegraphics[width=1.25in]{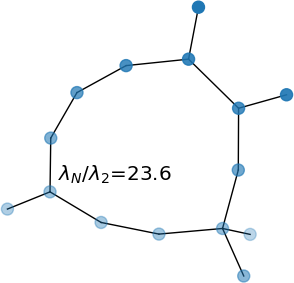}&
\includegraphics[width=1.25in]{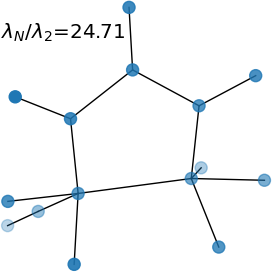}\\
$|Aut(Q)|=10$& $|Aut(Q)|=6$& $|Aut(Q)|=1$ &$|Aut(Q)|=2$\\
(a) & (b) & (c) & (d)\\
\end{tabular}
\caption{A collection of true ESB for the $3^{\textrm{rd}}$ generation of the $(1,2)$-flower graph, i.e. $F_3(1,2)$. If we quantify the ``amount of symmetry" by the size of the automorphism group of the network quotient graph~\cite{Xiao08}, then (a), being a minimal eigenratio ESB, has the highest degree of symmetry with $|Aut(Q)|=10$ with an eigenratio of $R= 21.79$; (b) shows a more quasi-symmetric ESB with $R=22.59$, where the action of using the network quotient vastly reduces the measure of symmetry $|Aut(Q)|=6$; (c) the ESB resulting from our greedy search (true ESB in this case) has $R=23.6$, and finally (d) (obtained from an exhaustive search for the ESB with largest eigenratio) has $R=24.71$, nearing the limiting value of $R_{MSF}\approx 24.8$. \label{MinimalESB}}
\end{figure}

We have also noted that the minimal eigenratio ESB, though not specifically preferred by our definition, do show indications that are relevant to forming intuition. We find that out of all small order ESB, those with minimal eigenratios are the ones with higher degrees of symmetry~\cite{Taylor20}, e.g., see the ESB of $F_3(1,2)$ provided in Fig.~\ref{Gallery}. This is likely due to the degeneracy that forms in the eigenspace; which has been explored through the nullity of the adjacency matrix $A$~\cite{Cheng07,Li08, Gong12}. This is indicative of the multiplicity of mid-range eigenvalues in $L$. However, it is nontrivial to measure symmetry meaningfully with respect to the process of synchronization. The automorphism group of a graph structure may include a large number of trivial symmetries with respect to this process, e.g., twin pendant swaps. Thus, we choose to measure ``large-scale" symmetry of the graph by the order of the automorphism group of the network quotient graph as described in \textit{Xiao et. al. (2008)}~\cite{Xiao08}, where redundant structures such as twin pendants are consolidated. Figure~\ref{MinimalESB} shows (a) the minimal ESB along with (b)-(d) three additional (non-minimal) ESB of the $F_3(1,2)$ graph. Although not the whole story, the general trend shows the synchronizability eigenratio $R=\lambda_N/\lambda_2$ increases as the order of the automorphism group of the network quotient, denoted here by $|Q|$, decreases. This trend is not expected to fully characterize the process of synchronization, but our analysis provides support for its relevance.

To more quantitatively explore this angle, it may be beneficial to consider the connectivity being defined by the random walk Laplacian $L_{rw}=\mathbb{I}_N-D^{-1}A$ instead of the combinatorial Laplacian (as is the general approach). Using a normalized Laplacian (though maintaining the rows summing to zero) may have interesting applications as well, where each node has a limited ability to adapt to external stimulus, meaning their response is based on the average stimulus and not the volume of stimulus. In this case, the research with respect to the nullity of $A$ would correspond directly to the multiplicity of $\lambda=1$ in $L_{rw}$, which would effectively contract the spectrum toward the center of the interval of values ($[0,2]$).

\section{Conclusions}
We have introduced a new concept associated with the structure interacting with dynamics on networks called the Essential Synchronization Backbone Problem (ESBP). The idea is that a core part of the graph, a subgraph, supports the synchrony and its associated information flow. We have explored solutions to this problem for two types of linear diffusively coupled chaotic oscillator systems, providing a greedy algorithm for finding approximate solutions, which we call Greedy Essential Synchronization Backbones (GESB). A true ESB can be found for small and sparse enough graphs, however, due to computational complexity issues, this approach is not tenable for general complex networks. The initial exploration of the solution space to this problem opens many directions of related research, including optimal information transfer and the role of hierarchy and graph conductance (or isoperimetric number) in the synchronization of oscillators.  

\section{Acknowledgments}
CTD has recieved funding from the Air Force Office of Scientific Research (20RICOR010)

EB has received funding from the Army Research Office (N68164-EG) and also DARPA
\section{Data Availability}
Code and data used to create figures available at \url{http://github.com/tylerdiggans/ESBP}.

\appendix
\section{Chaotic Attractors}
\label{Chaotics}
We introduce two chaotic attractors, which are used as the uncoupled dynamics of the networks oscillators systems.
\begin{enumerate}[(a)]
\item \textbf{The Lorenz Attractor}

Initially introduced as a simplified model for convection in weather patterns, the \textit{Lorenz system} is a three-dimensional dynamical system that can be described by a first order nonlinear system of differential equations. This simple model, having only two nonlinear (quadratic) terms results in quite an array of interesting phenomena. The state variables $\textbf{x}=[x,y,z]$, are governed by the equations of motion
\begin{equation}
\begin{array}{l}
\dot{x} = \sigma (y - x)\\
\dot{y} = x (\rho - z) - y\\
\dot{z} = xy - \beta z
\end{array},
\end{equation}
where $\sigma, \rho$ and $\beta$ are parameters, most commonly taken to be $\sigma=10$, $\rho=28$ and $\beta=8/3$ to induce chaos.

\vskip 0.1in
\item \textbf{The R\"{o}ssler Attractor}

The \textit{R\"{o}ssler system} provides a more permissive contrast to assess the effects of network topology with changing dynamics. This system has a single nonlinear term in its equations of motion, and when projected into the $z=0$ plane, the R\"{o}ssler system follows a linear differential equation. It is only through the introduction of the $z$ direction that chaos emerges. The same state variables $\textbf{x}=[x,y,z]$, are governed by
\begin{equation}
\begin{array}{l}
\dot{x} = - y - z\\
\dot{y} = x + a y\\
\dot{z} = b + z (x-c)
\end{array},
\end{equation}
where $a$, $b$, and $c$ are parameters. We will use $a=b=0.2$ and $c=9.0$ throughout. 

\end{enumerate} 

\begin{figure}[ht!]
\centering
\begin{tabular}{ccc}
\includegraphics[width=1.5in]{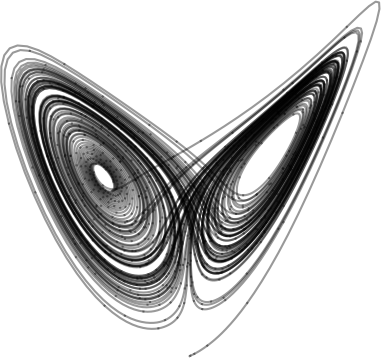} & \qquad\qquad\qquad \qquad \qquad\qquad&
\includegraphics[width=1.25in]{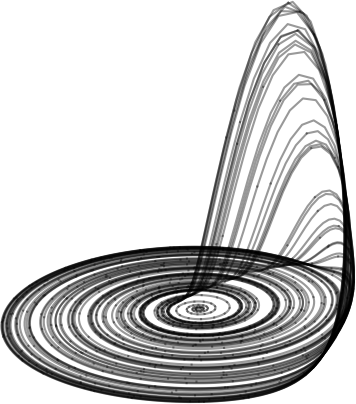}\\
(a) && (b)\\
\end{tabular}
\caption{An example trajectory of (a) the Lorenz attractor, and (b) the R\"{o}ssler attractor.\label{Oscillators}}
\end{figure}

\section{Master Stability Function}
\label{MSF}
The stability of the synchronization manifold for a large class of coupled dynamical systems was addressed by the introduction of the Master Stability Function~\cite{Pecora98} (MSF) for analysis of systems of the form:
\begin{equation*}
\dot{\vec{X}}=F(\vec{X})- \sigma L\otimes H(\vec{X}),
\end{equation*} 
as either the dynamical equations themselves or as the local linearization of nonlinearly coupled systems. In either case, it enabled the separation of the influence of the network topology from that of the oscillator dynamics and coupling on the stability of the synchronous state. 

The MSF is formed by considering the variational equation
\begin{equation}
\dot{\xi} = [\textbf{1}_N \otimes Df - \sigma L\otimes DH]\xi,
\label{Variation}
\end{equation}
where the $i$-th component of $\xi$, represents the variation of the variables at node $i$. Utilizing the eigen-decomposition of the connectivity matrix, $L$, we obtain a block diagonal form for Eq.~(\ref{Variation}). Given a particular choice of $f$ and $\hat{H}$ as described in section~\ref{Background}, the MSF is defined to be a function $\Psi:\mathbb{R}\rightarrow\mathbb{R}$, whose output is the maximum Lyapunov (or Floquet) exponent of the generic variational equation 
\begin{equation}
\dot{\zeta} = [Df- K \hat{H}] \zeta.
\end{equation}
The independent variable, $K$, plays the role of a coupling-normalized eigenvalue, e.g. $K=\sigma\lambda$, where $\lambda$ is an eigenvalue of the graph Laplacian. The smallest eigenvalue of $L$ for a synchronizable system is $\lambda_1=0$ and it relates to the trajectory of the synchronous state in systems of identical oscillators, since the coupling terms go to zero. Since $\Psi(K)$ is defined to be the largest Lyapunov exponent, whenever $\Psi(K)<0$ we are assured that all transverse directions are stable, meaning small perturbations in any of these directions will return to the synchronization manifold. Thus, for a specific global coupling constant $\sigma$, if $\Psi(\sigma\lambda_k)<0$ for $k=2,...,N$, then we have an exponentially asymptotically stable synchronization manifold.It is important to point out that not all systems will be amenable to MSF analysis as it has been described here, e.g., systems coupled through directed network with high degrees of non-normality~\cite{Fish17,Muolo21}.

\begin{figure}[ht!]
\centering
\begin{tabular}{ccc}
\includegraphics[width= 2.5in]{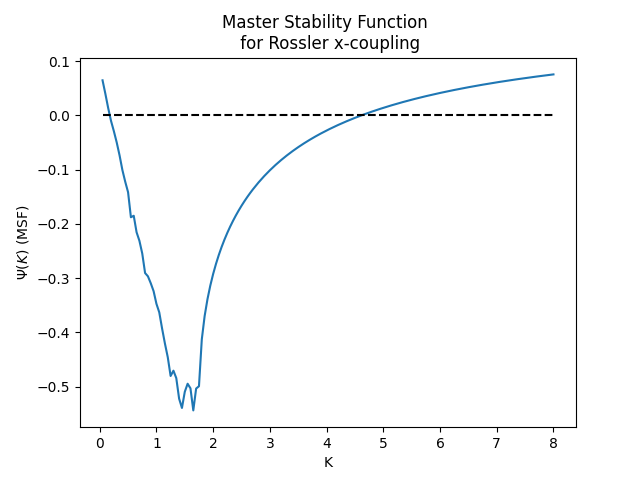}&
\qquad&
\includegraphics[width=2.5in]{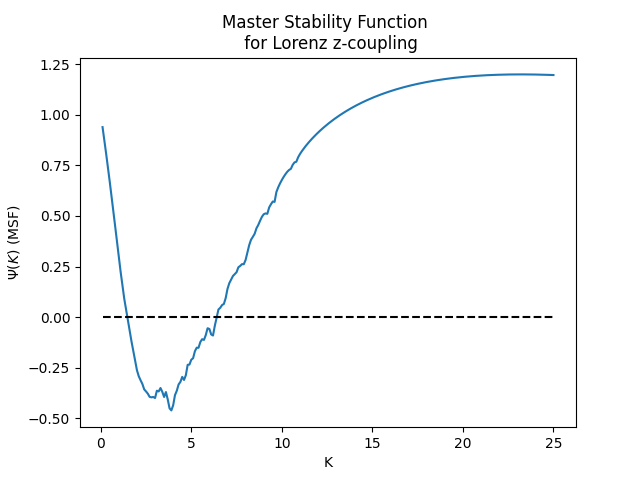}\\
(a) && (b)\\
\end{tabular}
\caption{Numerical approximations of the MSF $\left(\Psi(K)\right)$ for two systems consisting of identical dynamical systems at $N$ nodes with an associated coupling matrix $\hat{H}$: (a) R\"{o}ssler attractors coupled through $\hat{H}_x$ and (b) Lorenz attractors coupled through $\hat{H}_z$. All approximations were made using the $QR$ method~\cite{Bremen97} following guidance in \textit{Huang et. al., (2009)}~\cite{Huang09}.\label{MSFs}}
\end{figure}

A comprehensive study of most well-known chaotic systems has been carried out~\cite{Huang09}, and we provide the relevant plots of $\Psi$ in Fig.~\ref{MSFs}, which were recreated for the two choices considered in this paper: R\"{o}ssler oscillators using $\hat{H}_x$ and Lorenz oscillators using $\hat{H}_z$.  

A common measure of synchronizability for a network has been quantified by the eigenratio $R=\lambda_N/\lambda_2$ through direct comparison to the synchronizability ratio $R_{MSF}=K_\beta/K_\alpha$, where $K_\beta$ may be infinite. If $R<R_{MSF}$, then there exists a global coupling strength $\sigma$ such that~(\ref{x}) has a linearly stable synchronization manifold; furthermore, the smaller the $R$, the larger the range of global coupling strengths that enable such synchronization.

\bibliographystyle{unsrt}
\bibliography{ESBP_Bib}

\end{document}